\documentclass[aps,prl,twocolumn,floatfix,reprint]{revtex4-1}
\usepackage{graphicx}
\usepackage{color}
\usepackage{amssymb} 
\usepackage{amsmath} 
\usepackage{natbib}

\begin{document}

\title{Valence state manipulation of single Fe impurities in GaAs by STM}

\author{J. Bocquel$^1$}
\author{V.R. Kortan$^2$}
\author{R.P. Campion$^3$}
\author{B.L. Gallagher$^3$}
\author{M.E. Flatt\'e$^2$}
\author{P.M. Koenraad$^1$}

\affiliation{{$^1$\,Department of Applied Physics, Eindhoven University of Technology, P.O. Box 513, 5600 MB, Eindhoven, The Netherlands} \\
{$^2$\,Optical Science and Technology Center and Department of Physics and Astronomy, University of Iowa, Iowa City, Iowa 52242, USA}\\
{$^3$\,School of Physics and Astronomy, University of Nottingham, Nottingham NG7 2RD, United Kingdom}}

\begin{abstract}
The incorporation of Fe in GaAs was studied by cross-sectional scanning tunneling microscopy (X-STM).  The observed local electronic contrast of a single Fe atom is found to depend strongly on its charge state. We demonstrate that an applied tip voltage can be used to manipulate the valence and spin state of single Fe impurities in GaAs. In particular we can induce a transition from the (Fe$^{3+}$)$^0$ - 3d$^5$ - isoelectronic state to the (Fe$^{2+}$)$^-$ - 3d$^6$ - ionized acceptor state with an associated change of the spin moment.  Fe atoms sometimes produce dark anisotropic features in topographic maps, which is consistent with an interference between different tunneling paths.

\end{abstract}

\maketitle
The properties of single impurities in semiconductor hosts are of great fundamental and practical importance. As the dimensions of semiconductor devices become increasingly smaller, single impurities have an increasingly important influence on the device characteristics. This is problematic for conventional device designs but may make possible new ``solotronic" devices where a single impurity determines the functionality of the complete device \cite{Koenraad2011}. Furthermore the incorporation and electronic structure of magnetic impurities in semiconductor hosts are fundamental to the understanding of diluted magnetic semiconductors \cite{Dietl2000}.

Cross-sectional scanning tunneling microscopy (X-STM) provides the atomic-scale electronic resolution required to illuminate many of the fundamental properties of single impurities in semiconductor materials. In this work we use X-STM to investigate Fe doped GaAs at the atomic scale and report on the spatial distribution of Fe atoms and the different charge and valence states of substitutional Fe. We manipulate the valence state, {\it i.e.} the core spin of the Fe, with the tip voltage from (Fe$^{3+}$)$^0-$3d$^5$ to (Fe$^{2+}$)$^--$3d$^6$, where the first symbols describe the valence ($sp$) electrons and the second describe the core spin $d$-shell occupancy. This manipulation differs from the ionization of   Mn  in GaAs, [(Mn$^{2+}$)$^--$3d$^5$, h$^+$]$^0$ to (Mn$^{2+}$)$^--$3d$^5$ plus a free hole \cite{Yakunin2004}, where  ``h$^+$"  indicates the hole bound in a host-like acceptor state. The neutral [(Fe$^{2+}$)$^--$3d$^6$, h$^+$]$^0$ acceptor state of Fe atoms at a p-type surface has been observed, but not previously manipulated \cite{Richardella2009}. Change of the $d$-state occupation of Fe should strongly modify the magnetic moment of the Fe atom, of value to single impurity spintronics.

The X-STM measurements were performed both at room temperature and 77~K under UHV conditions (5$\times$10$^{-11}$ Torr). Electrochemically etched tungsten STM tips were used. The STM was operated in constant current mode on a clean and atomically flat GaAs (110) surface obtained by \textsl{in-situ} cleavage. The molecular beam epitaxy grown sample contains a 100 nm Fe doped GaAs layer (nominal concentration of $2\times 10^{18}$ cm$^{-3}$ ) and a Fe monolayer incorporated in GaAs. The growth temperature was 580$^\circ$C during the entire growth procedure. The nominal layer structure consisted of GaAs substrate/100nm Fe:GaAs/200nm GaAs/Fe monolayer/500nm GaAs. The growth was monitored by RHEED. In this paper we only report on the Fe monolayer region of this sample.

\begin{figure}[t]
\centering
\includegraphics[width=\columnwidth]{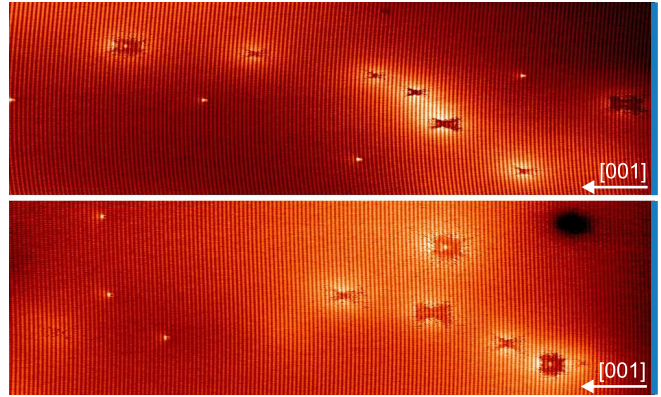}
\caption{
\small
90~nm $\times$ 35~nm filled states topography  of the segregation of Fe atoms in GaAs. The blue line at far right indicates the nominal position of the Fe monolayer.}
\label{segregation}
\end{figure}

In Fig.~\ref{segregation} two typical images of the GaAs layer directly above the Fe monolayer are shown. Every single feature which is dark and anisotropic is the signature of a single Fe atom. The nominal position of the Fe monolayer is indicated by the blue line at the right edge. No Fe atoms are detected at this position; instead  they have strongly segregated along the growth direction. Segregation has already been reported for other transition metals in III-V semiconductors. In the case of  Mn $\delta$-doped layers in GaAs a strong segregation occurs and SIMS \cite{Nazmul2003}  and X-STM investigations \cite{Bozkurt2010} have demonstrated segregation over distances of several tens of nm, which depends on the growth temperature and scales with the nominal Mn concentration. Here Fe atoms are found as far as 150 nm from the intended position of the Fe monolayer. However, no clear Fe concentration gradient is observed and an  accurate doping profile could not be determined due to poor statistics. This inhomogeneous and slow incorporation of the Fe atoms indicates that Fe acts as a surfactant during the growth of the GaAs capping layer. Such behavior is consistent with the low solubility reported for Fe in GaAs. An alternative explanation could be the Fermi-level pinning of the growth surface. The presence of an electric field at the surface can lead to the redistribution of the impurities during growth \cite{Luftman1990}.

\begin{figure}[t]
\centering
\includegraphics[width=\columnwidth]{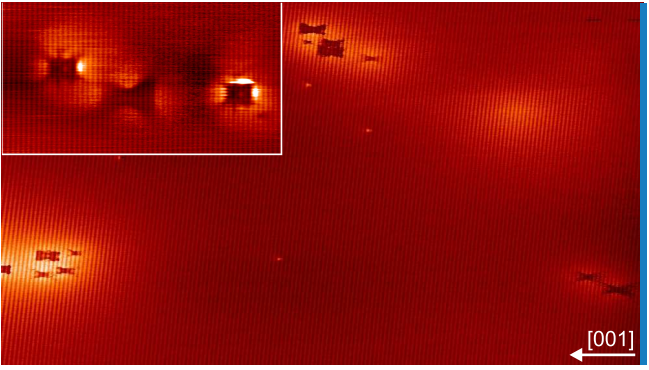}
\caption{
\small
90~nm $\times$ 50~nm filled states topography  showing dilute Fe clusters in GaAs. The clusters are found on a brighter background due to Coulomb interactions induced by the ionized Fe. The blue line indicates the nominal position of the Fe monolayer. The inset shows a close up of three Fe impurities.}
\label{clusters}
\end{figure}

The segregated Fe atoms in the capping layer are not distributed randomly. In Fig. \ref{clusters}, regions with high concentrations of Fe are apparent. These Fe-rich regions consist of Fe atoms that group together during epitaxy. The change in RHEED patterns from 2D to 3D after the growth of the Fe monolayer agrees with this picture. No nano-precipitates or secondary crystallographic phases are detected. It is important to stress that the Fe atoms are not in nearest neighbor positions but form dilute Fe clusters which do not seem to influence the crystal structure of the semiconductor host. The tendency of transition metals to aggregate during the epitaxial growth of III-V semiconductors has already been established \cite{Sato2010}. The mechanisms behind this are long range attractive forces between magnetic atoms, due to Coulombic interaction or strain field interactions. The local density fluctuations we observe resemble the so-called spinodal decomposition observed in some dilute magnetic semiconductors \cite{Sato2005,Katayama-Yoshida2007}, except that from our study there seems to be an absence of Fe-Fe bonds. From our X-STM 2D cuts, it seems that these isolated dilute clusters form extended three dimensional structures.

\begin{figure}[t]
\centering
\includegraphics[width=\columnwidth]{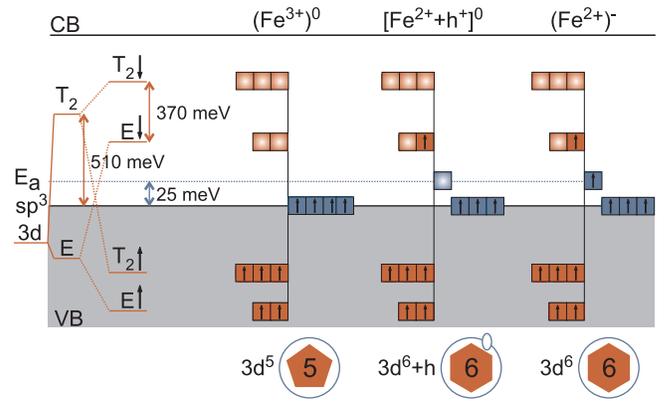}
\caption{Possible electronic configurations of an Fe atom in GaAs. The $3d$ electrons of Fe populate localized impurity-like levels (orange) mainly located in the valence band, as well as a delocalized anti-bonding state (blue) arising from their hybridization with As neighbors. The charge transfer level from Fe$^{2+}$ (5 $d$-electrons) to Fe$^{3+}$ (6 $d$-electrons) correspond to a localized energy level found 510~meV above the VB edge. The energy level associated to the bound hole state (Fe$^{2+}$)$^-$+ h$^{+}$ is found 25~meV above the VB edge. The most stable configuration is determined by the position of the Fermi level relative to these energy levels.}
\label{energylevels}
\end{figure}

The known possible electronic configurations \cite{Malguth2008} of an Fe atom in GaAs are illustrated in~Fig. \ref{energylevels}. Like most 3d transition metals, Fe impurities in III-V semiconductors act as charge traps and recombination centers, pinning the Fermi level at a deep energy level and inducing semi-insulating properties within the host. These deep levels, so-called charge transfer (CT) levels, correspond to a change in the valence state of the Fe atom and are expected to have a localized atomic-like character. The stable electronic configuration is determined by the position of the Fermi level in the semiconductor relative to the CT levels. Thus in GaAs, when substituting for the Ga cation site, an Fe atom can be in two different valence states: Fe$^{3+}$ or Fe$^{2+}$. Fe$^{3+}$ has a half filled $d$-shell and acts as an isoelectronic impurity, while Fe$^{2+}$ with 6 electrons in the $d$-shell acts as an acceptor. Fe will be in its acceptor state or in its isoelectronic state depending on whether the Fermi level is found above or below the Fe$^{2+}$/Fe$^{3+}$ CT level which is located 510~meV above the valence band edge \cite{Malguth2008}. Fe in its acceptor state Fe$^{2+}$ can be in two different charge states, A$^{0}$ or A$^{-}$. A$^{-}$ corresponds to the ionized acceptor (Fe$^{2+}$)$^{-}$ and A$^{0}$ to the neutral acceptor state [Fe$^{2+}$, h$^{+}$]$^{0}$. This neutral state corresponds to a hole bound to the ionized  acceptor (Fe$^{2+}$)$^-$ carrying a single negative charge \cite{Malguth2008, Dornen1993a}, and is expected to have a delocalized, host-like character. The ionization energy associated to these charge states, {\it i.e.} the energy needed for (Fe$^{2+}$)$^-$ to bind a free hole and form the [Fe$^{2+}$, h$^{+}$]$^{0}$ complex, is 25~meV. The three electronic configurations reported for Fe impurities in GaAs and described in Fig. \ref{energylevels} can be denoted by the following: (Fe$^{3+}$)$^0-$3d$^5$, [Fe$^{2+}$, h$^+$]$^0-$3d$^6$ + h$^+$ and (Fe$^{2+}$)$^--$3d$^6$.

\begin{figure}[t]
\centering
\includegraphics[width=\columnwidth]{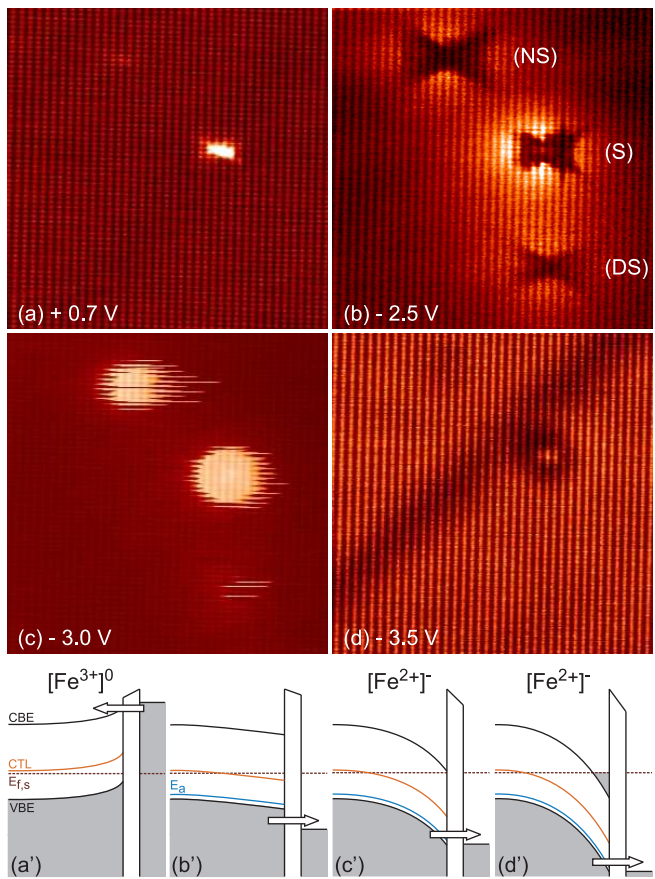}
\caption{
\small
(a-d)23~nm $\times$ 23~nm topography images of single Fe impurities in GaAs taken at a unique position but different voltages and their corresponding band alignements (a'-d'). Bringing the W tip close to the sample surface induces a bending of the semiconductor bands and shifts the associated energy levels. Changing the voltage between tip and sample results in changing locally the position of the Fermi level in the bandgap. As a consequence, the Fe impurities are brought into their different charge states. 
(a) isoelectronic acceptor Fe$^{3+}$. (b) and (c) ionized acceptor (Fe$^{2+}$)$^-$. (d) ionized acceptor (Fe$^{2+}$)$^-$ screened by an accumulation of electrons at the surface. The Fe impurities are located at different positions from the cleaved surface: in the surface layer (S), near the surface (NS) and deep below the surface (DS).}
\label{v_dependence}
\end{figure}

Figure~\ref{v_dependence} shows constant current filled states topography  taken at different voltages at a position where three Fe atoms are visible. The local electronic contrast induced by each Fe impurity is strongly dependent on the applied voltage. We distinguish four regimes for which Fe atoms give rise to images with distinctly different contrast and shapes due to applied voltage dependent local manipulation of the Fermi level position due to tip-induced band bending (TIBB). The Fermi level position within the band gap controls the electronic configuration of Fe atoms in the semiconductor.  Thus, tuning the voltage is expected to change the charge state and/or the valence of the impurities as described in Fig. \ref{v_dependence}.  Impurity charge manipulation using an STM tip or a nearby charged defect has already been demonstrated for various impurities in III-V semiconductors \cite{Yakunin2003,Teichmann2008,Lee2011}.

At sufficient positive voltage, some Fe atoms appear as bright highly localized features while the majority are not visible. Locally under the tip the semiconductor's bands as well as the CT level are pushed upwards at positive voltages and the sample Fermi level will be located below the  Fe$^{2+}$/Fe$^{3+}$ CT level. Under these conditions the Fe atoms are expected to be in their isoelectronic configuration Fe$^{3+}$. The contrast observed for isoelectronic Fe will originate from the difference in LDOS of the Fe atoms and the Ga atoms. This gives rise to a highly localized contrast so only impurities located in the surface layer (S) are expected to appear in the STM images. The contrast shown in Fig. \ref{v_dependence}(a) is therefore consistent with Fe atoms in the Fe$^{3+}$ configuration. The slightly delocalized features can be understood as a consequence of a $p-d$ hybridization of Fe with the valence band wavefunctions.

Fig. \ref{v_dependence}(c) shows that for applied voltage of $\approx$ -3.0 volts the Fe atoms appear as bright discs. Such discs correspond to the ionization of either donors \cite{Teichmann2008} or acceptors \cite{Lee2011}. In these studies, the onset of the disc indicates the spatial separation between the tip and the impurity results in a local TIBB  sufficient to ionize the impurity. 
The isotropic Coulomb field created by the ionized impurity induces a local enhancement of the states available for tunneling, and thus the tunnel current. For tunneling conditions like Fig.~\ref{v_dependence}(c), the Fe impurities are in their isoelectronic configuration when the tip is far away. Bringing the tip closer, at negative voltage, bends the bands and energy levels downwards. When the Fe$^{2+}$/Fe$^{3+}$CT level crosses the semiconductor Fermi level, the most favorable valence state for Fe switches from Fe$^{3+}$ to (Fe$^{2+}$)$^-$. The novelty in our system is that the Fe impurities release a core $d$-state hole, while former studies on Mn acceptors and Si donors report respectively on the loss of a bound valence hole or a bound conduction electron. Thus the neutral state here is Fe$^{3+}$,  not the effective-mass-like complex [Fe$^{2+}$, h$^{+}$]$^{0}$. The filling of the $3d$-shell changes from five to six electrons, and therefore the core spin state changes from $S=5/2$ to $S=2$. The presence of these bright discs in our topography images is a clear manifestation of the manipulation of the $d$-states of single Fe impurities by the STM tip.

Although, as with Mn acceptors and Si donors, the diameter of the discs increases with the applied voltage and tip-sample distance, there were additional features  detected which could be specific to a change in $d$-shell occupancy. The position of the disc edge depends on the nature of the atoms under the tip; a larger diameter occurs when the tip is located above the As sub-lattice. This could originate from stronger hybridization between the $d$-states of Fe and the host valence band states. Moreover,  an instability is observed  for some Fe impurities when the local Fermi level is very close to the CT level, due to spontaneous relaxation from (Fe$^{2+}$)$^-$ to Fe$^{3+}$. Analogous behaviour was recently described for Si atoms at the (110) surface of GaAs \cite{Garleff2011}.

Figure \ref{v_dependence}(d) shows the electronic contrast appearing  for Fe atoms at very large negative voltage. The semiconductor's bands are bent further downwards such that an accumulation of electrons is created at the surface of the semiconductor. These electrons screen charged defects and lead to charge density oscillations around each negatively charged (Fe$^{2+}$)$^-$ ion. The presence of such Friedel-like oscillations has already been reported for ionized dopants such as Si\cite{VanderWielenMC1996} and Mn\cite{Kitchen2007} in GaAs. To observe this requires a sample Fermi level relatively high in the band gap; for Fe-doped GaAs, the Fermi level is close to the CT level, located deep in the band gap.

Figure \ref{v_dependence}(b) shows that at low negative bias voltage, the electronic contrast in the topography  due to a single Fe atom is very peculiar, with all Fe atoms appearing as dark anisotropic features on a brighter background.  The shape of these features depends on the depth of the Fe atom from the surface. Moreover, the shape of a single impurity changes rapidly with the voltage and the current setpoint. From very low to moderate negative voltage, the dark feature becomes smaller and more asymmetric before disappearing. It shrinks from a large cubic shape to a small triangular shape through a bow tie and a cross feature. The same dependence is found when decreasing the current setpoint.

Similar anisotropic shapes (but with bright contrast) have been reported for the wave functions of holes bound to individual acceptors in GaAs \cite{Yakunin2004,Mahieu2005,Richardella2009}. Bound hole wave functions are imaged at low positive voltage, when electrons can tunnel from the tip to the empty acceptor state and appear as bright anisotropic features. The anisotropy of the contrast originates from the cubic symmetry of the host crystal \cite{Yakunin2004}, while the asymmetry is related to the interaction with the surface and the binding energy of the acceptor \cite{Celebi2010}. Thus, the shape variation of these features from one Fe atom to the other is attributed to the depth dependence of the charge distribution contrast, as shown for Mn impurities in GaAs \cite{Garleff2008}. The voltage dependence is understood as a consequence of the TIBB. The shapes become smaller and more asymmetric as the TIBB becomes larger. The current dependence is weaker and can not be understood as a result of a change in TIBB. The asymmetry is stronger when the current setpoint is low, i.e when the tip-sample distance is large and the TIBB smaller. We think that the shape variation with the current setpoint is induced by the difference in the overlap of the tip and the sample wavefunctions.  Decreases in conductivity appearing at a single impurity in filled states topography images are rare. Standard electron tunneling mechanisms from a  semiconductor sample to a metallic tip can not explain our observations and the origin of this dark contrast around the Fe impurity remains unclear.

\begin{figure}[t]
\centering
\includegraphics[width=\columnwidth]{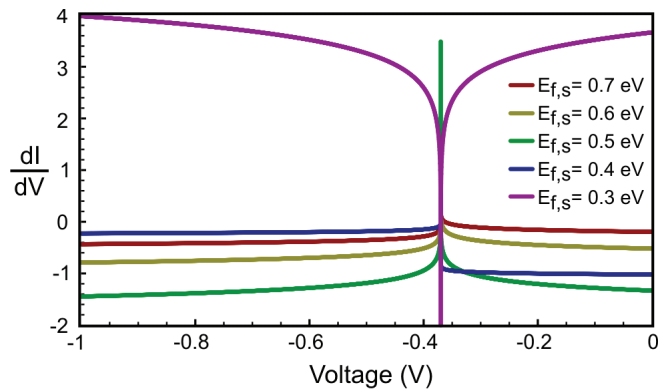}
\caption{
\small
Calculated contribution to dI/dV from the tunneling path that virtually excites a core exciton, in units of the dI/dV from the other tunneling path (that does not),  for several potential values of the Fermi energy E$_{f,s}$.  near the impurity. Parameters and the quantitative expression (after Ref.~\onlinecite{Persson1987}) used are described in supplementary information. Negative values indicate a suppression of the tunneling current.}
\label{calculs}
\end{figure}

We suggest a possible interpretation which draws on observations of a decrease in tunneling current from interference between two elastic tunneling paths through a single molecule, one path virtually exciting a vibrational resonance \cite{Davis1970,Bayman1981,Persson1987,Baratoff1988}. Here we consider a similar scenario in which one path virtually excites an electronic transition, from $E$ to $T_2$ crystal-field-split states in the $d$ shell of the Fe impurity (a core exciton). Using similar formalism as Ref.~\onlinecite{Persson1987}, and described in more detail in the Supplemental Material \cite{SupMat}, we find the $dI/dV$ curves shown in Fig.~\ref{calculs}. The large regions of negative calculated $\partial I/ \partial V$, which would appear dark in topographic maps relative to surroundings, suggest an interpretation for the dark anisotropic shapes shown in Fig.~4(b), and indicate that interference effects could be the cause of the decrease in tunneling current for negative applied voltages.

In summary we studied the spatial distribution and the electronic state of Fe in GaAs by cross-sectional scanning tunneling microscopy (X-STM). We achieved the manipulation of the core state of single Fe impurities in GaAs by controlling the TIBB, via the applied voltage. Fe atoms can be brought from their isoelectronic state (Fe$^{3+}$)$^0-$3d$^5$, into an ionized acceptor state (Fe$^{2+}$)$^--$3d$^6$. The observed local electronic contrast of a single Fe atom was found to change strongly due to different charge states. Finally, a peculiar contrast was observed, where Fe atoms appear as dark anisotropic features, that we tentatively explain by an interference between two elastic tunneling paths.

We  acknowledge funding from the European Community's Seventh Framework Programme (PF7/2007-2013) SemiSpinNet.


\begin{thebibliography}{26}%
\makeatletter
\providecommand \@ifxundefined [1]{%
 \@ifx{#1\undefined}
}%
\providecommand \@ifnum [1]{%
 \ifnum #1\expandafter \@firstoftwo
 \else \expandafter \@secondoftwo
 \fi
}%
\providecommand \@ifx [1]{%
 \ifx #1\expandafter \@firstoftwo
 \else \expandafter \@secondoftwo
 \fi
}%
\providecommand \natexlab [1]{#1}%
\providecommand \enquote  [1]{``#1''}%
\providecommand \bibnamefont  [1]{#1}%
\providecommand \bibfnamefont [1]{#1}%
\providecommand \citenamefont [1]{#1}%
\providecommand \href@noop [0]{\@secondoftwo}%
\providecommand \href [0]{\begingroup \@sanitize@url \@href}%
\providecommand \@href[1]{\@@startlink{#1}\@@href}%
\providecommand \@@href[1]{\endgroup#1\@@endlink}%
\providecommand \@sanitize@url [0]{\catcode `\\12\catcode `\$12\catcode
  `\&12\catcode `\#12\catcode `\^12\catcode `\_12\catcode `\%12\relax}%
\providecommand \@@startlink[1]{}%
\providecommand \@@endlink[0]{}%
\providecommand \url  [0]{\begingroup\@sanitize@url \@url }%
\providecommand \@url [1]{\endgroup\@href {#1}{\urlprefix }}%
\providecommand \urlprefix  [0]{URL }%
\providecommand \Eprint [0]{\href }%
\providecommand \doibase [0]{http://dx.doi.org/}%
\providecommand \selectlanguage [0]{\@gobble}%
\providecommand \bibinfo  [0]{\@secondoftwo}%
\providecommand \bibfield  [0]{\@secondoftwo}%
\providecommand \translation [1]{[#1]}%
\providecommand \BibitemOpen [0]{}%
\providecommand \bibitemStop [0]{}%
\providecommand \bibitemNoStop [0]{.\EOS\space}%
\providecommand \EOS [0]{\spacefactor3000\relax}%
\providecommand \BibitemShut  [1]{\csname bibitem#1\endcsname}%
\let\auto@bib@innerbib\@empty
\bibitem [{\citenamefont {Koenraad}\ and\ \citenamefont
  {Flatt\'{e}}(2011)}]{Koenraad2011}%
  \BibitemOpen
  \bibfield  {author} {\bibinfo {author} {\bibfnamefont {P.~M.}\ \bibnamefont
  {Koenraad}}\ and\ \bibinfo {author} {\bibfnamefont {M.~E.}\ \bibnamefont
  {Flatt\'{e}}},\ }\href {\doibase 10.1038/nmat2940} {\bibfield  {journal}
  {\bibinfo  {journal} {Nature Materials}\ }\textbf {\bibinfo {volume} {10}},\
  \bibinfo {pages} {91} (\bibinfo {year} {2011})}\BibitemShut {NoStop}%
\bibitem [{\citenamefont {Dietl}\ \emph {et~al.}(2000)\citenamefont {Dietl},
  \citenamefont {Ohno}, \citenamefont {Matsukura}, \citenamefont {Cibert},\
  and\ \citenamefont {Ferrand}}]{Dietl2000}%
  \BibitemOpen
  \bibfield  {author} {\bibinfo {author} {\bibfnamefont {T.}~\bibnamefont
  {Dietl}}, \bibinfo {author} {\bibfnamefont {H.}~\bibnamefont {Ohno}},
  \bibinfo {author} {\bibfnamefont {F.}~\bibnamefont {Matsukura}}, \bibinfo
  {author} {\bibfnamefont {J.}~\bibnamefont {Cibert}}, \ and\ \bibinfo {author}
  {\bibfnamefont {D.}~\bibnamefont {Ferrand}},\ }\href@noop {} {\bibfield
  {journal} {\bibinfo  {journal} {Science}\ }\textbf {\bibinfo {volume} {287}}
  (\bibinfo {year} {2000})}\BibitemShut {NoStop}%
\bibitem [{\citenamefont {Yakunin}\ \emph {et~al.}(2004)\citenamefont
  {Yakunin}, \citenamefont {Silov}, \citenamefont {Koenraad}, \citenamefont
  {Wolter}, \citenamefont {{Van Roy}}, \citenamefont {{De Boeck}},
  \citenamefont {Tang},\ and\ \citenamefont {Flatt\'{e}}}]{Yakunin2004}%
  \BibitemOpen
  \bibfield  {author} {\bibinfo {author} {\bibfnamefont {A.~M.}~\bibnamefont
  {Yakunin}}, \bibinfo {author} {\bibfnamefont {A.~Y.}~\bibnamefont {Silov}},
  \bibinfo {author} {\bibfnamefont {P.~M.}~\bibnamefont {Koenraad}}, \bibinfo
  {author} {\bibfnamefont {J.~H.}~\bibnamefont {Wolter}}, \bibinfo {author}
  {\bibfnamefont {W.}~\bibnamefont {{Van Roy}}}, \bibinfo {author}
  {\bibfnamefont {J.}~\bibnamefont {{De Boeck}}}, \bibinfo {author}
  {\bibfnamefont {J.-M.}\ \bibnamefont {Tang}}, \ and\ \bibinfo {author}
  {\bibfnamefont {M.~E.}~\bibnamefont {Flatt\'{e}}},\ }\href {\doibase
  10.1103/PhysRevLett.92.216806} {\bibfield  {journal} {\bibinfo  {journal}
  {Physical Review Letters}\ }\textbf {\bibinfo {volume} {92}},\ \bibinfo
  {pages} {216806} (\bibinfo {year} {2004})}\BibitemShut {NoStop}%
\bibitem [{\citenamefont {Richardella}\ \emph {et~al.}(2009)\citenamefont
  {Richardella}, \citenamefont {Kitchen},\ and\ \citenamefont
  {Yazdani}}]{Richardella2009}%
  \BibitemOpen
  \bibfield  {author} {\bibinfo {author} {\bibfnamefont {A.}~\bibnamefont
  {Richardella}}, \bibinfo {author} {\bibfnamefont {D.}~\bibnamefont
  {Kitchen}}, \ and\ \bibinfo {author} {\bibfnamefont {A.}~\bibnamefont
  {Yazdani}},\ }\href {\doibase 10.1103/PhysRevB.80.045318} {\bibfield
  {journal} {\bibinfo  {journal} {Physical Review B}\ }\textbf {\bibinfo
  {volume} {80}},\ \bibinfo {pages} {045318} (\bibinfo {year}
  {2009})}\BibitemShut {NoStop}%
\bibitem [{\citenamefont {Nazmul}\ \emph {et~al.}(2003)\citenamefont {Nazmul},
  \citenamefont {Sugahara},\ and\ \citenamefont {Tanaka}}]{Nazmul2003}%
  \BibitemOpen
  \bibfield  {author} {\bibinfo {author} {\bibfnamefont {A.}~\bibnamefont
  {Nazmul}}, \bibinfo {author} {\bibfnamefont {S.}~\bibnamefont {Sugahara}}, \
  and\ \bibinfo {author} {\bibfnamefont {M.}~\bibnamefont {Tanaka}},\ }\href
  {\doibase 10.1016/S0022-0248(02)02274-1} {\bibfield  {journal} {\bibinfo
  {journal} {Journal of Crystal Growth}\ }\textbf {\bibinfo {volume} {251}},\
  \bibinfo {pages} {303} (\bibinfo {year} {2003})}\BibitemShut {NoStop}%
\bibitem [{\citenamefont {Bozkurt}\ \emph {et~al.}(2010)\citenamefont
  {Bozkurt}, \citenamefont {Grant}, \citenamefont {Ulloa}, \citenamefont
  {Campion}, \citenamefont {Foxon}, \citenamefont {Marega}, \citenamefont
  {Salamo},\ and\ \citenamefont {Koenraad}}]{Bozkurt2010}%
  \BibitemOpen
  \bibfield  {author} {\bibinfo {author} {\bibfnamefont {M.}~\bibnamefont
  {Bozkurt}}, \bibinfo {author} {\bibfnamefont {V.~A.}\ \bibnamefont {Grant}},
  \bibinfo {author} {\bibfnamefont {J.~M.}\ \bibnamefont {Ulloa}}, \bibinfo
  {author} {\bibfnamefont {R.~P.}\ \bibnamefont {Campion}}, \bibinfo {author}
  {\bibfnamefont {C.~T.}\ \bibnamefont {Foxon}}, \bibinfo {author}
  {\bibfnamefont {E.}~\bibnamefont {Marega}}, \bibinfo {author} {\bibfnamefont
  {G.~J.}\ \bibnamefont {Salamo}}, \ and\ \bibinfo {author} {\bibfnamefont
  {P.~M.}\ \bibnamefont {Koenraad}},\ }\href {\doibase 10.1063/1.3293296}
  {\bibfield  {journal} {\bibinfo  {journal} {Applied Physics Letters}\
  }\textbf {\bibinfo {volume} {96}},\ \bibinfo {pages} {042108} (\bibinfo
  {year} {2010})}\BibitemShut {NoStop}%
\bibitem [{\citenamefont {Schubert}\ \emph {et~al.}(1990)\citenamefont
  {Schubert}, \citenamefont {Kuo}, \citenamefont {Kopf}, \citenamefont
  {Jordan}, \citenamefont {Luftman},\ and\ \citenamefont
  {Hopkins}}]{Luftman1990}%
  \BibitemOpen
  \bibfield  {author} {\bibinfo {author} {\bibfnamefont {E.~F.}~\bibnamefont
  {Schubert}}, \bibinfo {author} {\bibfnamefont {J.~M.}~\bibnamefont {Kuo}},
  \bibinfo {author} {\bibfnamefont {R.~F.}~\bibnamefont {Kopf}}, \bibinfo {author}
  {\bibfnamefont {A.~S.}~\bibnamefont {Jordan}}, \bibinfo {author} {\bibfnamefont
  {H.~S.}~\bibnamefont {Luftman}}, \ and\ \bibinfo {author} {\bibfnamefont
  {L.~C.}~\bibnamefont {Hopkins}},\ }\href@noop {} {\bibfield  {journal} {\bibinfo
   {journal} {Physical Review B}\ }\textbf {\bibinfo {volume} {42}},\ \bibinfo
  {pages} {1364} (\bibinfo {year} {1990})}\BibitemShut {NoStop}%
\bibitem [{\citenamefont {Sato}\ \emph {et~al.}(2010)\citenamefont {Sato},
  \citenamefont {Kudrnovsk\'{y}}, \citenamefont {Dederichs}, \citenamefont
  {Eriksson}, \citenamefont {Turek}, \citenamefont {Sanyal}, \citenamefont
  {Bouzerar}, \citenamefont {Katayama-Yoshida}, \citenamefont {Dinh},
  \citenamefont {Fukushima}, \citenamefont {Kizaki},\ and\ \citenamefont
  {Zeller}}]{Sato2010}%
  \BibitemOpen
  \bibfield  {author} {\bibinfo {author} {\bibfnamefont {K.}~\bibnamefont
  {Sato}}, \bibinfo {author} {\bibfnamefont {J.}~\bibnamefont
  {Kudrnovsk\'{y}}}, \bibinfo {author} {\bibfnamefont {P.~H.}\ \bibnamefont
  {Dederichs}}, \bibinfo {author} {\bibfnamefont {O.}~\bibnamefont {Eriksson}},
  \bibinfo {author} {\bibfnamefont {I.}~\bibnamefont {Turek}}, \bibinfo
  {author} {\bibfnamefont {B.}~\bibnamefont {Sanyal}}, \bibinfo {author}
  {\bibfnamefont {G.}~\bibnamefont {Bouzerar}}, \bibinfo {author}
  {\bibfnamefont {H.}~\bibnamefont {Katayama-Yoshida}}, \bibinfo {author}
  {\bibfnamefont {V.~A.}\ \bibnamefont {Dinh}}, \bibinfo {author}
  {\bibfnamefont {T.}~\bibnamefont {Fukushima}}, \bibinfo {author}
  {\bibfnamefont {H.}~\bibnamefont {Kizaki}}, \ and\ \bibinfo {author}
  {\bibfnamefont {R.}~\bibnamefont {Zeller}},\ }\href {\doibase
  10.1103/RevModPhys.82.1633} {\bibfield  {journal} {\bibinfo  {journal}
  {Reviews of Modern Physics}\ }\textbf {\bibinfo {volume} {82}},\ \bibinfo
  {pages} {1633} (\bibinfo {year} {2010})}\BibitemShut {NoStop}%
\bibitem [{\citenamefont {Sato}\ \emph {et~al.}(2005)\citenamefont {Sato},
  \citenamefont {Katayama-Yoshida},\ and\ \citenamefont
  {Dederichs}}]{Sato2005}%
  \BibitemOpen
  \bibfield  {author} {\bibinfo {author} {\bibfnamefont {K.}~\bibnamefont
  {Sato}}, \bibinfo {author} {\bibfnamefont {H.}~\bibnamefont
  {Katayama-Yoshida}}, \ and\ \bibinfo {author} {\bibfnamefont {P.~H.}\
  \bibnamefont {Dederichs}},\ }\href {\doibase 10.1143/JJAP.44.L948} {\bibfield
   {journal} {\bibinfo  {journal} {Japanese Journal of Applied Physics}\
  }\textbf {\bibinfo {volume} {44}},\ \bibinfo {pages} {L948} (\bibinfo {year}
  {2005})}\BibitemShut {NoStop}%
\bibitem [{\citenamefont {Katayama-Yoshida}\ \emph {et~al.}(2007)\citenamefont
  {Katayama-Yoshida}, \citenamefont {Sato}, \citenamefont {Fukushima},
  \citenamefont {Toyoda}, \citenamefont {Kizaki}, \citenamefont {Dinh},\ and\
  \citenamefont {Dederichs}}]{Katayama-Yoshida2007}%
  \BibitemOpen
  \bibfield  {author} {\bibinfo {author} {\bibfnamefont {H.}~\bibnamefont
  {Katayama-Yoshida}}, \bibinfo {author} {\bibfnamefont {K.}~\bibnamefont
  {Sato}}, \bibinfo {author} {\bibfnamefont {T.}~\bibnamefont {Fukushima}},
  \bibinfo {author} {\bibfnamefont {M.}~\bibnamefont {Toyoda}}, \bibinfo
  {author} {\bibfnamefont {H.}~\bibnamefont {Kizaki}}, \bibinfo {author}
  {\bibfnamefont {V.~A.}\ \bibnamefont {Dinh}}, \ and\ \bibinfo {author}
  {\bibfnamefont {P.~H.}\ \bibnamefont {Dederichs}},\ }\href {\doibase
  10.1002/pssa.200673021} {\bibfield  {journal} {\bibinfo  {journal} {Physica
  Status Solidi (A)}\ }\textbf {\bibinfo {volume} {204}},\ \bibinfo {pages}
  {15} (\bibinfo {year} {2007})}\BibitemShut {NoStop}%
\bibitem [{\citenamefont {Malguth}\ \emph {et~al.}(2008)\citenamefont
  {Malguth}, \citenamefont {Hoffmann},\ and\ \citenamefont
  {Phillips}}]{Malguth2008}%
  \BibitemOpen
  \bibfield  {author} {\bibinfo {author} {\bibfnamefont {E.}~\bibnamefont
  {Malguth}}, \bibinfo {author} {\bibfnamefont {A.}~\bibnamefont {Hoffmann}}, \
  and\ \bibinfo {author} {\bibfnamefont {M.~R.}\ \bibnamefont {Phillips}},\
  }\href {\doibase 10.1002/pssb.200743315} {\bibfield  {journal} {\bibinfo
  {journal} {Physica Status Solidi (B)}\ }\textbf {\bibinfo {volume} {245}},\
  \bibinfo {pages} {455} (\bibinfo {year} {2008})}\BibitemShut {NoStop}%
\bibitem [{\citenamefont {Pressel}\ \emph {et~al.}(1993)\citenamefont
  {Pressel}, \citenamefont {Dornen},\citenamefont {Ruckert},\ and\ \citenamefont
  {Thonke}}]{Dornen1993a}%
  \BibitemOpen
  \bibfield  {author} {\bibinfo {author} {\bibfnamefont {K.}~\bibnamefont
  {Pressel}}, \bibinfo {author} {\bibfnamefont {A.}~\bibnamefont {Dornen}},	
	\bibinfo {author} {\bibfnamefont {G.}~\bibnamefont  {Ruckert}}, \ 
	and\ \bibinfo {author} {\bibfnamefont {K.}~\bibnamefont {Thonke}},\
  }\href@noop {} {\bibfield  {journal} {\bibinfo  {journal} {Physical Review
  B}\ }\textbf {\bibinfo {volume} {47}},\ \bibinfo {pages} {16267} (\bibinfo
  {year} {1993})}\BibitemShut {NoStop}%
\bibitem [{\citenamefont {Yakunin}\ \emph {et~al.}(2003)\citenamefont
  {Yakunin}, \citenamefont {Silov}, \citenamefont {Koenraad}, \citenamefont
  {{Van Roy}}, \citenamefont {{De Boeck}},\ and\ \citenamefont
  {Wolter}}]{Yakunin2003}%
  \BibitemOpen
  \bibfield  {author} {\bibinfo {author} {\bibfnamefont {A.~M.}~\bibnamefont
  {Yakunin}}, \bibinfo {author} {\bibfnamefont {A.~Y.}~\bibnamefont {Silov}},
  \bibinfo {author} {\bibfnamefont {P.~M.}\ \bibnamefont {Koenraad}}, \bibinfo
  {author} {\bibfnamefont {W.}~\bibnamefont {{Van Roy}}}, \bibinfo {author}
  {\bibfnamefont {J.}~\bibnamefont {{De Boeck}}}, \ and\ \bibinfo {author}
  {\bibfnamefont {J.~H.}\ \bibnamefont {Wolter}},\ }\href {\doibase
  10.1016/j.spmi.2004.03.055} {\bibfield  {journal} {\bibinfo  {journal}
  {Superlattices and Microstructures}\ }\textbf {\bibinfo {volume} {34}},\
  \bibinfo {pages} {539} (\bibinfo {year} {2003})}\BibitemShut {NoStop}%
\bibitem [{\citenamefont {Teichmann}\ \emph {et~al.}(2008)\citenamefont
  {Teichmann}, \citenamefont {Wenderoth}, \citenamefont {Loth}, \citenamefont
  {Ulbrich}, \citenamefont {Garleff}, \citenamefont {Wijnheijmer},\ and\
  \citenamefont {Koenraad}}]{Teichmann2008}%
  \BibitemOpen
  \bibfield  {author} {\bibinfo {author} {\bibfnamefont {K.}~\bibnamefont
  {Teichmann}}, \bibinfo {author} {\bibfnamefont {M.}~\bibnamefont
  {Wenderoth}}, \bibinfo {author} {\bibfnamefont {S.}~\bibnamefont {Loth}},
  \bibinfo {author} {\bibfnamefont {R.~G.}~\bibnamefont {Ulbrich}}, \bibinfo
  {author} {\bibfnamefont {J.~K.}~\bibnamefont {Garleff}}, \bibinfo {author}
  {\bibfnamefont {A.~P.}~\bibnamefont {Wijnheijmer}}, \ and\ \bibinfo {author}
  {\bibfnamefont {P.~M.}\ \bibnamefont {Koenraad}},\ }\href {\doibase
  10.1103/PhysRevLett.101.076103} {\bibfield  {journal} {\bibinfo  {journal}
  {Physical Review Letters}\ }\textbf {\bibinfo {volume} {101}},\ \bibinfo
  {pages} {076103} (\bibinfo {year} {2008})}\BibitemShut {NoStop}%
\bibitem [{\citenamefont {Lee}\ and\ \citenamefont {Gupta}(2011)}]{Lee2011}%
  \BibitemOpen
  \bibfield  {author} {\bibinfo {author} {\bibfnamefont {D.~H.}~\bibnamefont
  {Lee}}\ and\ \bibinfo {author} {\bibfnamefont {J.~A.}~\bibnamefont {Gupta}},\
  }\href {http://pubs.acs.org/doi/abs/10.1021/nl2003686} {\bibfield  {journal}
  {\bibinfo  {journal} {Nano Letters}\ }\textbf {\bibinfo {volume} {11}},\
  \bibinfo {pages} {2004} (\bibinfo {year} {2011})}\BibitemShut {NoStop}%
\bibitem [{\citenamefont {Garleff}\ \emph {et~al.}(2011)\citenamefont
  {Garleff}, \citenamefont {Wijnheijmer}, \citenamefont {v.~d. Enden},\ and\
  \citenamefont {Koenraad}}]{Garleff2011}%
  \BibitemOpen
  \bibfield  {author} {\bibinfo {author} {\bibfnamefont {J.~K.}~\bibnamefont
  {Garleff}}, \bibinfo {author} {\bibfnamefont {A.~P.}~\bibnamefont
  {Wijnheijmer}}, \bibinfo {author} {\bibfnamefont {C.~N.}\ \bibnamefont {v.~d.
  Enden}}, \ and\ \bibinfo {author} {\bibfnamefont {P.~M.}~\bibnamefont
  {Koenraad}},\ }\href {\doibase 10.1103/PhysRevB.84.075459} {\bibfield
  {journal} {\bibinfo  {journal} {Physical Review B}\ }\textbf {\bibinfo
  {volume} {84}},\ \bibinfo {pages} {075459} (\bibinfo {year}
  {2011})}\BibitemShut {NoStop}%
\bibitem [{\citenamefont {van~der Wielen}\ \emph {et~al.}(1996)\citenamefont
  {van~der Wielen}, \citenamefont {van Roij},\ and\ \citenamefont {van
  Kempen}}]{VanderWielenMC1996}%
  \BibitemOpen
  \bibfield  {author} {\bibinfo {author} {\bibfnamefont {M.~C.~M.~M.}~\bibnamefont
  {van~der Wielen}}, \bibinfo {author} {\bibfnamefont {A.~J.~A.}~\bibnamefont {van
  Roij}}, \ and\ \bibinfo {author} {\bibfnamefont {H.}~\bibnamefont {van
  Kempen}},\ }\href {http://www.ncbi.nlm.nih.gov/pubmed/10061627} {\bibfield
  {journal} {\bibinfo  {journal} {Physical Review Letters}\ }\textbf {\bibinfo
  {volume} {76}},\ \bibinfo {pages} {1075} (\bibinfo {year}
  {1996})}\BibitemShut {NoStop}%
\bibitem [{\citenamefont {Kitchen}\ \emph {et~al.}(2007)\citenamefont
  {Kitchen}, \citenamefont {Richardella}, \citenamefont {Roushan},
  \citenamefont {Tang}, \citenamefont {Flatt\'{e}},\ and\ \citenamefont
  {Yazdani}}]{Kitchen2007}%
  \BibitemOpen
  \bibfield  {author} {\bibinfo {author} {\bibfnamefont {D.}~\bibnamefont
  {Kitchen}}, \bibinfo {author} {\bibfnamefont {A.}~\bibnamefont
  {Richardella}}, \bibinfo {author} {\bibfnamefont {P.}~\bibnamefont
  {Roushan}}, \bibinfo {author} {\bibfnamefont {J.-M.}\ \bibnamefont {Tang}},
  \bibinfo {author} {\bibfnamefont {M.~E.}~\bibnamefont {Flatt\'{e}}}, \ and\
  \bibinfo {author} {\bibfnamefont {A.}~\bibnamefont {Yazdani}},\ }\href
  {\doibase 10.1063/1.2694511} {\bibfield  {journal} {\bibinfo  {journal}
  {Journal of Applied Physics}\ }\textbf {\bibinfo {volume} {101}},\ \bibinfo
  {pages} {09G515} (\bibinfo {year} {2007})}\BibitemShut {NoStop}%
\bibitem [{\citenamefont {Mahieu}\ \emph {et~al.}(2005)\citenamefont {Mahieu},
  \citenamefont {Grandidier}, \citenamefont {Deresmes}, \citenamefont {Nys},
  \citenamefont {Sti\'{e}venard},\ and\ \citenamefont {Ebert}}]{Mahieu2005}%
  \BibitemOpen
  \bibfield  {author} {\bibinfo {author} {\bibfnamefont {G.}~\bibnamefont
  {Mahieu}}, \bibinfo {author} {\bibfnamefont {B.}~\bibnamefont {Grandidier}},
  \bibinfo {author} {\bibfnamefont {D.}~\bibnamefont {Deresmes}}, \bibinfo
  {author} {\bibfnamefont {J.~P.}~\bibnamefont {Nys}}, \bibinfo {author}
  {\bibfnamefont {D.}~\bibnamefont {Sti\'{e}venard}}, \ and\ \bibinfo {author}
  {\bibfnamefont {P.}~\bibnamefont {Ebert}},\ }\href {\doibase
  10.1103/PhysRevLett.94.026407} {\bibfield  {journal} {\bibinfo  {journal}
  {Physical Review Letters}\ }\textbf {\bibinfo {volume} {94}},\ \bibinfo
  {pages} {026407} (\bibinfo {year} {2005})}\BibitemShut {NoStop}%
\bibitem [{\citenamefont {\c{C}elebi}\ \emph {et~al.}(2010)\citenamefont
  {\c{C}elebi}, \citenamefont {Garleff}, \citenamefont {Silov}, \citenamefont
  {Yakunin}, \citenamefont {Koenraad}, \citenamefont {VanRoy}, \citenamefont
  {Tang},\ and\ \citenamefont {Flatt\'{e}}}]{Celebi2010}%
  \BibitemOpen
  \bibfield  {author} {\bibinfo {author} {\bibfnamefont {C.}~\bibnamefont
  {\c{C}elebi}}, \bibinfo {author} {\bibfnamefont {J.~K.}\ \bibnamefont
  {Garleff}}, \bibinfo {author} {\bibfnamefont {A.~Y.}~\bibnamefont {Silov}},
  \bibinfo {author} {\bibfnamefont {A.~M.}~\bibnamefont {Yakunin}}, \bibinfo
  {author} {\bibfnamefont {P.~M.}\ \bibnamefont {Koenraad}}, \bibinfo {author}
  {\bibfnamefont {W.}~\bibnamefont {VanRoy}}, \bibinfo {author} {\bibfnamefont
  {J.-M.}\ \bibnamefont {Tang}}, \ and\ \bibinfo {author} {\bibfnamefont
  {M.~E.}\ \bibnamefont {Flatt\'{e}}},\ }\href {\doibase
  10.1103/PhysRevLett.104.086404} {\bibfield  {journal} {\bibinfo  {journal}
  {Physical Review Letters}\ }\textbf {\bibinfo {volume} {104}},\ \bibinfo
  {pages} {086404} (\bibinfo {year} {2010})}\BibitemShut {NoStop}%
\bibitem [{\citenamefont {Garleff}\ \emph {et~al.}(2008)\citenamefont
  {Garleff}, \citenamefont {\c{C}elebi}, \citenamefont {{Van Roy}},
  \citenamefont {Tang}, \citenamefont {Flatt\'{e}},\ and\ \citenamefont
  {Koenraad}}]{Garleff2008}%
  \BibitemOpen
  \bibfield  {author} {\bibinfo {author} {\bibfnamefont {J.~K.}~\bibnamefont
  {Garleff}}, \bibinfo {author} {\bibfnamefont {C.}~\bibnamefont {\c{C}elebi}},
  \bibinfo {author} {\bibfnamefont {W.}~\bibnamefont {{Van Roy}}}, \bibinfo
  {author} {\bibfnamefont {J.-M.}\ \bibnamefont {Tang}}, \bibinfo {author}
  {\bibfnamefont {M.~E.}~\bibnamefont {Flatt\'{e}}}, \ and\ \bibinfo {author}
  {\bibfnamefont {P.~M}~\bibnamefont {Koenraad}},\ }\href {\doibase
  10.1103/PhysRevB.78.075313} {\bibfield  {journal} {\bibinfo  {journal}
  {Physical Review B}\ }\textbf {\bibinfo {volume} {78}},\ \bibinfo {pages}
  {075313} (\bibinfo {year} {2008})}\BibitemShut {NoStop}%
\bibitem [{\citenamefont {Persson}\ and\ \citenamefont
  {Baratoff}(1987)}]{Persson1987}%
  \BibitemOpen
  \bibfield  {author} {\bibinfo {author} {\bibfnamefont {B.~N.~J.}~\bibnamefont
  {Persson}}\ and\ \bibinfo {author} {\bibfnamefont {A.}~\bibnamefont
  {Baratoff}},\ }\href@noop {} {\bibfield  {journal} {\bibinfo  {journal}
  {Physical Review Letters}\ }\textbf {\bibinfo {volume} {59}},\ \bibinfo
  {pages} {339} (\bibinfo {year} {1987})}\BibitemShut {NoStop}%
\bibitem [{\citenamefont {Davis}(1970)}]{Davis1970}%
  \BibitemOpen
  \bibfield  {author} {\bibinfo {author} {\bibfnamefont {L.~C.}~\bibnamefont
  {Davis}},\ }\href@noop {} {\bibfield  {journal} {\bibinfo  {journal}
  {Physical Review B}\ }\textbf {\bibinfo {volume} {2}},\ \bibinfo {pages}
  {1714} (\bibinfo {year} {1970})}\BibitemShut {NoStop}%
\bibitem [{\citenamefont {Bayman}\ \emph {et~al.}(1981)\citenamefont {Bayman},
  \citenamefont {Hansma},\ and\ \citenamefont {Kaska}}]{Bayman1981}%
  \BibitemOpen
  \bibfield  {author} {\bibinfo {author} {\bibfnamefont {A.}~\bibnamefont
  {Bayman}}, \bibinfo {author} {\bibfnamefont {P.~K.}~\bibnamefont {Hansma}}, \
  and\ \bibinfo {author} {\bibfnamefont {W.~C.}~\bibnamefont {Kaska}},\
  }\href@noop {} {\bibfield  {journal} {\bibinfo  {journal} {Physical Review
  B}\ }\textbf {\bibinfo {volume} {24}},\ \bibinfo {pages} {2449} (\bibinfo
  {year} {1981})}\BibitemShut {NoStop}%
\bibitem [{\citenamefont {Baratoff}\ and\ \citenamefont
  {Persson}(1988)}]{Baratoff1988}%
  \BibitemOpen
  \bibfield  {author} {\bibinfo {author} {\bibfnamefont {A.}~\bibnamefont
  {Baratoff}}\ and\ \bibinfo {author} {\bibfnamefont {B.~N.~J.}~\bibnamefont
  {Persson}},\ }\href@noop {} {\bibfield  {journal} {\bibinfo  {journal}
  {Journal of Vacuum Science \& Technology A}\ }\textbf {\bibinfo {volume}
  {6}},\ \bibinfo {pages} {331} (\bibinfo {year} {1988})}\BibitemShut {NoStop}%
\bibitem [{Sup()}]{SupMat}%
  \BibitemOpen
  \href@noop {} {\bibinfo  {journal} {See Supplemental Material at [URL will be
  inserted by publisher] for the interference between elastic paths method}\
  }\BibitemShut {NoStop}%
\end{thebibliography}

%

\end{document}